\documentclass[aps, showpacs, showkeys,nofootinbib,floatfix]{revtex4}

\usepackage{amssymb}
\usepackage{amsmath}
\usepackage{graphicx}

\begin{document}

\title{A unification of RDE model and XCDM model }

\author{Kai Liao, Zong-Hong Zhu}
 \email{zhuzh@bnu.edu.cn}
\affiliation{Department of Astronomy, Beijing Normal University,
Beijing 100875, China}

\begin{abstract}
In this Letter, we propose a new generalized Ricci dark energy (NGR) model to unify Ricci dark energy (RDE)
and XCDM. Our model can distinguish between RDE and XCDM by introducing a parameter
$\beta$ called weight factor. When $\beta=1$, NGR model becomes the usual RDE model. The XCDM model is corresponding to $\beta=0$. Moreover,
NGR model permits the situation where neither $\beta=1$ nor $\beta=0$. We then perform a statefinder analysis on NGR model to see how $\beta$ effects the trajectory on the $r-s$ plane.
 In order to know the value of $\beta$, we constrain
NGR model with latest observations including type Ia supernovae (SNe Ia) from Union2 set (557 data), baryonic acoustic oscillation
(BAO) observation from the spectroscopic Sloan Digital Sky Survey (SDSS) data release 7 (DR7)
galaxy sample and cosmic microwave background (CMB) observation from the 7-year Wilkinson
Microwave Anisotropy Probe (WMAP7) results. With Markov Chain Monte
Carlo (MCMC) method, the constraint result is $\beta$=$0.08_{-0.21}^{+0.30}(1\sigma)_{-0.28}^{+0.43}(2\sigma)$,
which manifests the observations prefer a XCDM universe rather than RDE model. It seems RDE model is ruled out in NGR scenario within $2\sigma$ regions.
Furthermore, we compare it with some of successful cosmological models using AIC information criterion. NGR model seems to be a good
choice for describing the universe.

\end{abstract}


\maketitle

\section{$\text{Introduction}$}
Various cosmic observations suggest our universe is undergoing an accelerated expansion \cite{acceleration}. To explain this phenomenon, people introduce
an exotic component with negative pressure known as dark energy. The simplest dark energy model is cosmological constant ($\Lambda$)
\cite{lambda} or
XCDM model where dark energy has an arbitrary equation of state (EOS) $\omega_X$.
It fits all kinds of observational data well while it is confronted with theoretical problems such as "coincidence" problem and "fine-tuning" problem \cite{problem}. As a result, other dark energy models have been widely proposed including quintessence \cite{quintessence}, quintom \cite{quintom}, phantom \cite{phantom}, GCG \cite{GCG} and so on. In principle, dark energy is related to quantum gravity \cite{quantum}. But until now, a self-consistent quantum gravity theory has not established.
Nevertheless, the holographic principle \cite{principle} is thought to be a reflection of quantum gravity. Motivated by this, holographic dark energy
has been proposed. It embodies the relation between UV cut-off and IR cut off. However, how to choose the IR cut-off is a problem.
Cohen et al. \cite{first} first chose Hubble scale as IR cut-off. Hsu and Li \cite{point} pointed out it can not give an acceleration solution. Li then suggested the future event
horizon as IR cut-off \cite{FE}. Basing on causality, Cai proposed agegraphic dark energy \cite{AU} and new agegraphic dark energy \cite{CT}. Furthermore, Gao et al. \cite{RDE} proposed a holographic dark energy from Ricci scalar curvature. In RDE model, the IR cut-off is determined by a local quantity.

Nowadays, all the models above seem to be consistent with current observations. Usually, we estimate models through the $\chi^2$ or information criteria like BIC and AIC \cite{criteria}. In this Letter, we find XCDM model and RDE model can be related by a parameter $\beta$,
thus we can estimate them through constraining $\beta$. The distribution of $\beta$ can reflect which model is better. For example, if
the best-fit value of $\beta$ is close to 1 and 0 is not within 2-$\sigma$ range, we can say the observations support RDE model rather
than XCDM model. We now give some similar examples. In order to know whether $\Lambda$CDM is right, people free the EOS
parameter and constrain it with observations. If the result is close to -1, we can say $\Lambda$CDM is still a good choice. However, if the EOS parameter tends to -2, then $\Lambda$CDM should be suspected. Likewise, for purely dimensional reasons, Granda and Oliveros \cite{NIR} proposed a new IR cut-off. Wang and Xu \cite{NIR2} give the constraint results which suggest the coefficient of $H^2$ is two times larger than the one of $\dot{H}$, thus ruling out
the SRDE model \cite{SRDE}.
In RDE model, the density of dark energy is proportional to Ricci scalar or the sum of traces of energy-momentum tensors of each component. Since the trace of radiation is 0, we can ignore its impact on space-time curvature. RDE model suggests the weights of
dark energy and matter are the same, while XCDM model suggests only the trace of dark energy can affect its density. Therefore, what
on earth is the weight of matter (0.5 or 0?) is an interesting thing we want to know. Motivated by this, we free the weight of matter
as an arbitrary parameter called weight factor.

The rest of the Letter is organized as follows. In Section 2, we give the dynamics of the new generalized Ricci dark energy model. In Section 3, we give a statefinder diagnostic. In Section 4, we introduce the observational data we use. The constraint results are shown in Section 5. At last, we give the discussion and
conclusion in section 6. Throughout the Letter, the unit with light velocity $c=1$ is used.

\section{$\text{New generalized Ricci dark energy model}$}
We assume the universe is flat and described by Friedmann-Robertson-Walker (FRW) metric.
For RDE model, the density of dark energy is proportional to Ricci scalar
\begin{equation}
\rho_{de}\propto R.
\end{equation}

Considering Einstein field equation can be expressed as
\begin{equation}
R=8\pi G T,
\end{equation}
where G is Newtonian constant, T is the sum of traces of each component,
RDE model can be expressed as
\begin{equation}
\rho_{de}\propto T_{de}+T_m\propto \rho_{de}-3p+\rho_m.
\end{equation}

From this equation, the coefficients of $T_{de}$ and $T_m$ are both 1, which means the weights of dark energy and matter are the same.
We now change the weight of matter, the equation becomes
\begin{equation}
\rho_{de}=\alpha(T_{de}+\beta T_m)=\alpha(\rho_{de}-3p+\beta \rho_m),
\end{equation}
$\beta$ here is the weight factor we introduce which reflects the relative weight of matter to dark energy. If $\beta=1$, it becomes the
usual RDE model. When $\beta=0$
\begin{equation}
\rho_{de}=\alpha (\rho_{de}-3p),
\end{equation}
equivalently,
\begin{equation}
\rho_{de}\propto p,
\end{equation}
the NGR model becomes XCDM model.
For simplicity, we define dimensionless quantities $\Omega_m=\frac{\rho_m}{\rho_c}$,
$\Omega_{de}=\frac{\rho_{de}}{\rho_c}$, where $\rho_c=\frac{3H_0^2}{8\pi G}$ is the
critical density of the universe. $H$ is Hubble parameter, subscript "0" represents the
quantity today.
The Friedmann equation can be expressed as
\begin{equation}
E^2=\Omega_{de}+\Omega_m,
\end{equation}
where $E=\frac{H}{H_0}$.

The energy-momentum conservation equation can be expressed as
\begin{equation}
\dot{\Omega_{i}}+3H(1+\omega_i)\Omega_{i}=0,
\end{equation}
subscript "i" represents dark energy or matter.
Then we get
\begin{equation}
\Omega_{de}'=\frac{(4\alpha-1)\Omega_{de}+\alpha\beta\Omega_{m0}(1+z)^3}{\alpha(1+z)},
\end{equation}
where $\Omega_{de}'=\frac{d\Omega_{de}}{dz}$. With the initial condition
\begin{equation}
\Omega_{de0}+\Omega_{m0}=1,
\end{equation}
we can obtain the evolution of $\Omega_{de}$ with respect to redshift z
\begin{equation}
\Omega_{de}=(1-\frac{(\alpha\beta+1-\alpha)\Omega_{m0}}{1-\alpha})(1+z)^{4-\frac{1}{\alpha}}+\frac{\alpha\beta\Omega_{m0}(1+z)^3}{1-\alpha}.
\end{equation}

The EOS parameter can be obtained by
\begin{equation}
\omega_{de}=-1+(1+z)\frac{\Omega_{de}'}{3\Omega_{de}},
\end{equation}
and the deceleration parameter
\begin{equation}
q=\frac{1}{2}(1+\frac{3\omega_{de}\Omega_{de}}{\Omega_{de}+\Omega_{m}}).
\end{equation}

\begin{figure}[!htbp]
\includegraphics[width=80mm]{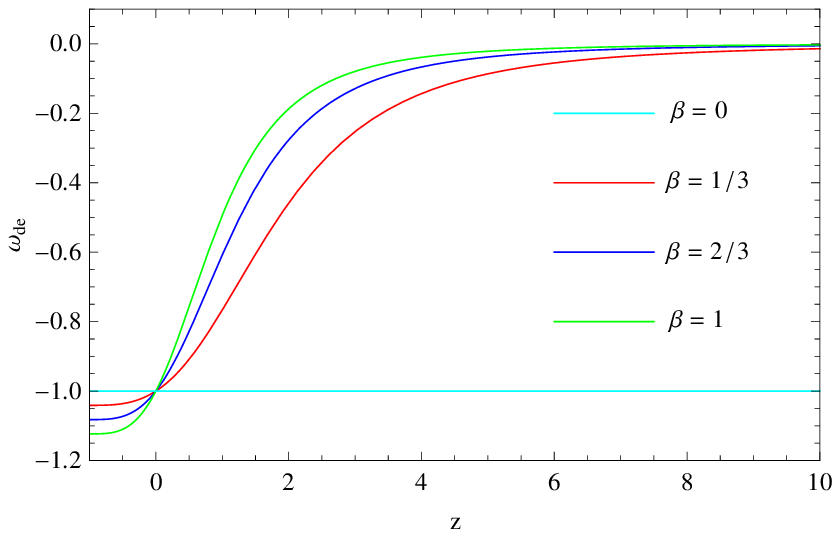}
\includegraphics[width=80mm]{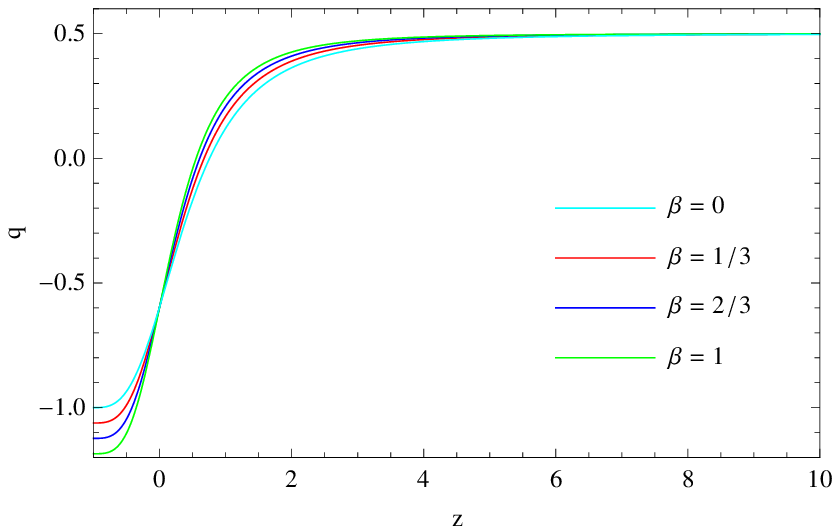}
\caption{The evolutions of $w_{de}(z)$ (left) and $q(z)$ (right) with respect to z in NGR model. $\Omega_{m0}=0.27$, $\omega_{de0}=-1$.
 }\label{fig1}
\end{figure}

\begin{figure}[!htbp]
\includegraphics[width=80mm]{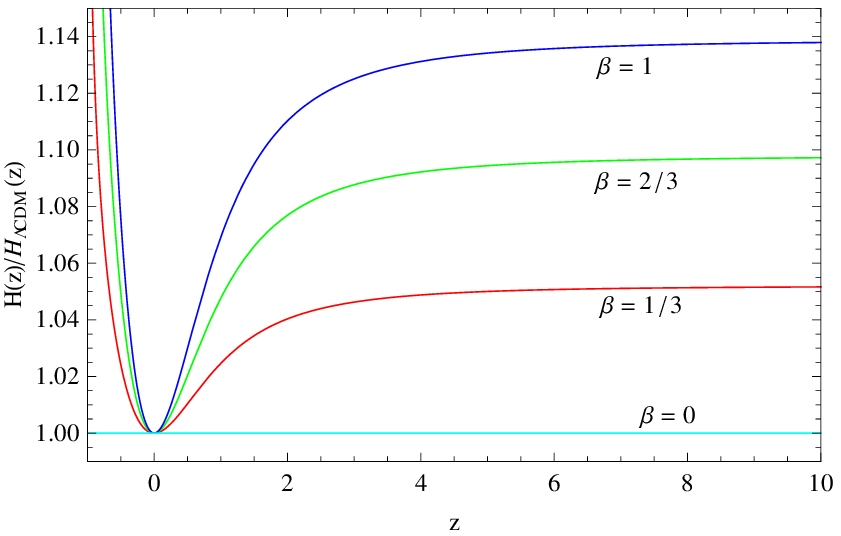}
\includegraphics[width=80mm]{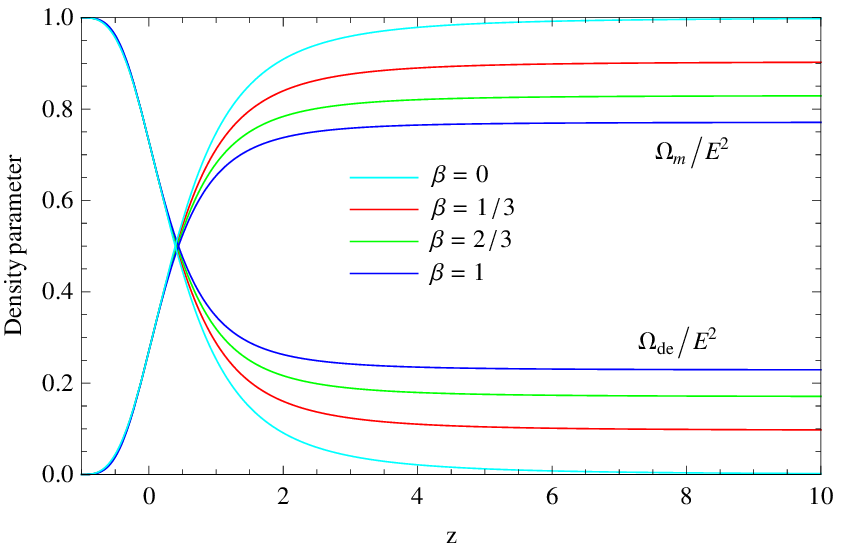}
\caption{The evolutions of the Hubble parameter in units of $H_{\Lambda CDM}(z)$ (left) and the density parameters (right). $\Omega_{m0}=0.27$, $\omega_{de0}=-1$.
 }\label{figh}
\end{figure}

In order to exhibit the effects of $\beta$, we fix $\Omega_{m0}=0.27$ and $\omega_{de0}=-1$ and plot the evolutions of $\omega_{de}(z)$,
$q(z)$, Hubble parameter $H(z)$ and density parameters defined as $\Omega_i/E^2$ in Fig. \ref{fig1} and Fig. \ref{figh}.

\begin{table*}
 \begin{center}
 \begin{tabular}{|c|c|c|c|} \hline\hline
 & \multicolumn{3}{c|}{The NGR Model}  \\
 \cline{2-4}                 &     $\Omega_{m0}$                  &     $\alpha$                       &     $\beta$ and $\chi_{\rm min}^2$         \\ \hline
$*$  \ \ & \ \ $0.284_{-0.035}^{+0.036}(1\sigma)_{-0.048}^{+0.050}(2\sigma)$ \ \  & \ \ $0.235_{-0.039}^{+0.046}(1\sigma)_{-0.053}^{+0.068}(2\sigma)$ \ \ & \ \
$0.08_{-0.21}^{+0.30}(1\sigma)_{-0.28}^{+0.43}(2\sigma)$   531.710 \ \ \\
$\beta=0$  \ \ & \ \ $0.280_{-0.029}^{+0.032}(1\sigma)_{-0.041}^{+0.050}(2\sigma)$ \ \  & \ \ $0.246_{-0.026}^{+0.029}(1\sigma)_{-0.037}^{+0.043}(2\sigma)$ \ \ & \ \ $*$  532.238\ \ \\
$\beta=0.5$  \ \ & \ \ $0.287_{-0.031}^{+0.034}(1\sigma)_{-0.044}^{+0.053}(2\sigma)$ \ \  & \ \ $0.195_{-0.019}^{+0.019}(1\sigma)_{-0.027}^{+0.029}(2\sigma)$ \ \ & \ \ $*$  539.734\ \ \\
$\beta=1$  \ \ & \ \ $0.296_{-0.033}^{+0.037}(1\sigma)_{-0.047}^{+0.054}(2\sigma)$ \ \  & \ \ $0.161_{-0.014}^{+0.016}(1\sigma)_{-0.021}^{+0.023}(2\sigma)$ \ \ & \ \ $*$  558.834\ \ \\

 \hline\hline

 \end{tabular}
 \end{center}\label{tab1}
 \caption{The best-fit values of  parameters and $\chi_{\rm min}^2$ for NGR model including the case where we fix $\beta=0.5$, as well as XCDM model and RDE model
 with the 1-$\sigma$ and 2-$\sigma$ uncertainties,
for the data sets SNe+BAO+CMB.
}\label{tab1}
 \end{table*}

\section{$\text{Statefinder diagnostic}$}
Statefinder diagnostic is a useful method to differentiate effective cosmological models since these models are all seen to be consist with current observations. It was first introduced by Sahni et al. \cite{statefinder}. This method probes the expansion dynamics of
the universe through high derivatives of scale factor $\dddot a$. The dimensionless statefinder pair $\{ r, s\}$ is defined as
\begin{equation}
    r\equiv\frac{\dddot a}{aH^3}, \quad s\equiv\frac{r-1}{3(q-1/2)} .
\end{equation}

Since the scale factor depends on the space-time manifold, the statefinder is a geometrical diagnostic. Different models are
corresponding to different trajectories on the $r-s$ plane. For example, the spatially flat $\Lambda$CDM model are corresponding
to a fixed point on the plane,
\begin{equation}
\{s,r\}\bigg\vert_{\rm \Lambda CDM} = \{ 0,1\} ~.\label{lcdm}
\end{equation}

Statefinder has been applied to various dark energy models including quintessence, quintom, GCG, braneworld model and so on \cite{applysf}.
We now turn to statefindr diagnostic for NGR model and find the effects of $\beta$. The statefinder parameters can also be expressed in terms of the total energy density and the total pressure
\begin{equation}
    r = 1 + \frac{9(\rho_{tot}+p)}{2\rho_{tot}}\frac{\dot p}{\dot\rho_{tot}} \,, \quad s = \frac{(\rho_{tot}+p)}{p}\frac{\dot p}{\dot\rho_{tot}}
    \, ,
\end{equation}
where we ignore the pressure of matter.

Combined with the dynamics we discussed in section 2, we have
\begin{equation}
s=\frac{\Omega_{m0}(1+z)^3+\frac{(1+z)\Omega_{de}'}{3}}{3(-\Omega_{de}+\frac{(1+z)\Omega_{de}'}{3})}\frac{3\beta\Omega_{m0}(1+z)^2+(1-\frac{1}{\alpha})\Omega_{de}'}{\Omega_{de}'+3\Omega_{m0}(1+z)^2},
\end{equation}
and
\begin{equation}
r=1+\frac{9}{2}\frac{-\Omega_{de}+\frac{(1+z)\Omega_{de}'}{3}}{\Omega_{de}+\Omega_{m0}(1+z)^3}s.
\end{equation}

In order to plot the statefinder plane, we fix the current EOS of dark energy and the density of matter as $\omega_{de0}=-1$ and
$\Omega_{m0}=0.27$, respectively.

\begin{figure}[!htbp]
\includegraphics[width=15cm]{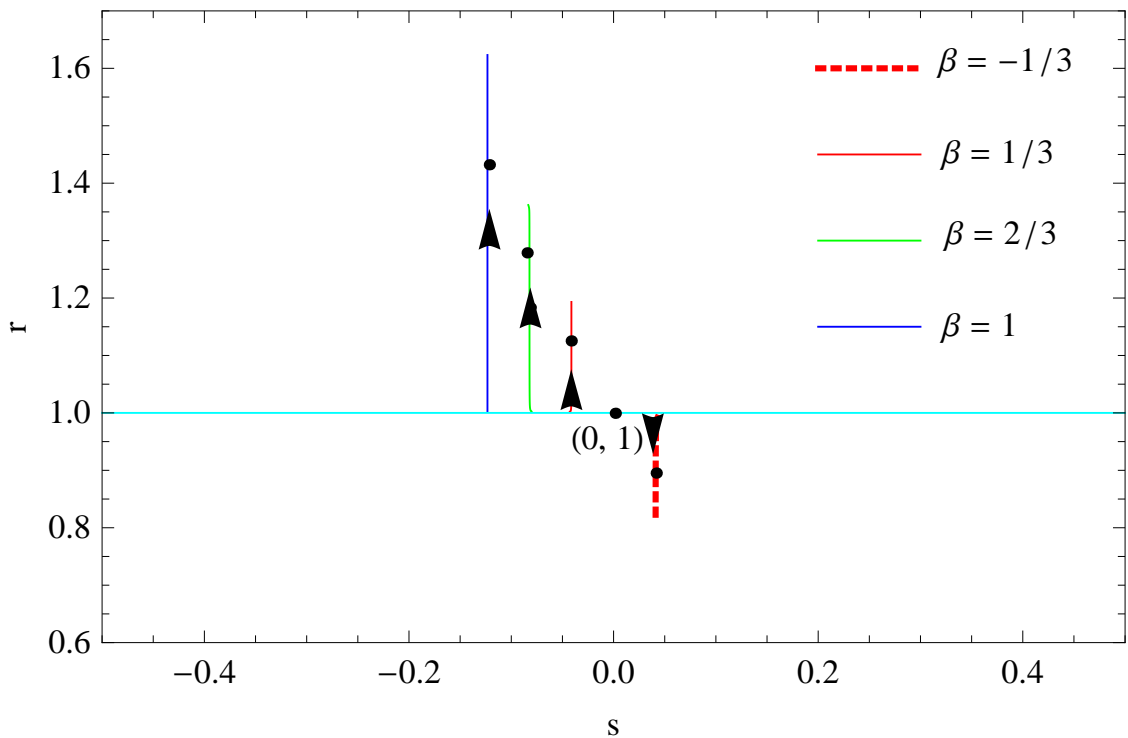}
\caption{The $r-s$ plane for NGR model with $\beta=-1/3, 0, 1/3, 2/3, 1$, respectively.
 }\label{figsf}
\end{figure}

In Fig. \ref{figsf}, we can see with the increase of the value of $\beta$, the corresponding s becomes smaller, and the range of the
trajectory becomes larger. For XCDM model, we choose the initial condition as $\omega_{de0}=-1$, it is regarded as $\Lambda$CDM model here. The dots represent the points today which are linear to $\beta$. r=0.865, 1.135, 1.27 and 1.405 for $\beta=-1/3, 1/3, 2/3, 1$, respectively. Our results are consist with
the RDE case \cite{sfRDE}.

\section{$\text{Current observational data}$}
\subsection{Type Ia supernovae}
SNe Ia has been an important tool for probing the nature of the universe since it first revealed the acceleration of the universe.
The current data (Union2) is given by the Supernova Cosmology Project (SCP)
collaboration including 557 samples \cite{Union2}. The distance modules can be expressed as
\begin{equation}
\mu=5 \log(d_L/\rm{Mpc})+25~,
\end{equation}
where $d_L$ is the luminosity distance. In a flat universe, it is related
to redshift which is a observational quantity
\begin{equation}
d_L={(1+z)} \int^{z}_0{dz'}/{H(z')}.
\end{equation}

We choose the marginalized nuisance parameter \cite{MNP} for
$\chi^2$:
\begin{equation}
\label{chi2SN} \chi^2_{\rm
SNe}=A-\frac{B^2}{C},
\end{equation}
where $A=\sum_i^{557}{(\mu^{\rm data}-\mu^{\rm
theory})^2}/{\sigma^2_{i}}$, $B=\sum_i^{557}{(\mu^{\rm
data}-\mu^{\rm theory})}/{\sigma^2_{i}}$, $C=\sum_i^{557}{1}/{\sigma^2_{i}}$,
$\sigma_{i}$ is the 1$\sigma$ uncertainty of the observational data.

\subsection{Baryon acoustic oscillation}
For BAO, the distance scale is expressed as \cite{DV}
\begin{equation} D_V(z)=\frac{1}{H_0}\big
[\frac{z}{E(z)}\big(\int_0^{z}\frac{dz}{E(z)}\big
)^2\big]^{1/3}~,
\end{equation}
and baryons were released from photons at the drag epoch. The corresponding redshift $z_d$ is
give by
\begin{equation}
z_{d}=\frac{1291(\Omega_{\mathrm{m0}}h^2)^{0.251}}{[1+0.659(\Omega_\mathrm{m0}h^2)^{0.828}]}[(1+b_{1}(\Omega_{b}h^2)^{b_2})],
\end{equation}
where
$b_1=0.313(\Omega_{\mathrm{m0}}h^2)^{-0.419}[1+0.607(\Omega_{\mathrm{m0}}h^2)^{0.674}]^{-1}$
and $b_2=0.238(\Omega_{\mathrm{m0}}h^2)^{0.223}$ \cite{BAOA}.
For BAO observation, we choose the measurements of the distance radio ($d_z$) at $z=0.2$ and $z=0.35$ \cite{BAOB}.
It can be defined as
\begin{equation}
d_z=\frac{r_s(z_d)}{D_V(z)},
\end{equation}
where $r_s(z_d)$ is the comoving sound horizon.
The SDSS data release 7 (DR7) galaxy sample gives the best-fit values
of the data set ($d_{0.2}$, $d_{0.35}$) \cite{BAOB}
\begin{eqnarray}
\hspace{-.5cm}\bar{\bf{P}}_{\rm matrix} &=& \left(\begin{array}{c}
{\bar d_{0.2}} \\
{\bar d_{0.35}}\\
\end{array}
  \right)=
  \left(\begin{array}{c}
0.1905\pm0.0061\\
0.1097\pm0.0036\\
\end{array}
  \right).
 \end{eqnarray}

The $\chi^2$ value of this BAO observation from SDSS DR7 can be
calculated as \cite{BAOB}
\begin{eqnarray}
\chi^2_{\rm BAO}=\Delta
\textbf{P}_{\rm matrix}^\mathrm{T}{\bf
C_{\rm matrix}}^{-1}\Delta\textbf{P}_{\rm matrix},
\end{eqnarray}
where $\Delta\bf{P_{\mathrm{matrix}}} =
\bf{P_{\mathrm{matrix}}}-\bf{\bar{P}_{\mathrm{matrix}}}$, and the
corresponding inverse  covariance matrix is
\begin{eqnarray}
\hspace{-.5cm} {\bf C_{\rm matrix}}^{-1}=\left(
\begin{array}{ccc}
30124& -17227\\
-17227& 86977\\
\end{array}
\right).
\end{eqnarray}

\subsection{Cosmic microwave background}
For CMB, the acoustic scale is related to the distance ratio. It can be
expressed as
\begin{equation}
l_a=\pi\frac{\Omega_\mathrm{k}^{-1/2}sinn[\Omega_\mathrm{k}^{1/2}\int_0^{z_{\ast}}\frac{dz}{E(z)}]/H_0}{r_s(z_{\ast})},
\end{equation}
where $r_s(z_{\ast})
={H_0}^{-1}\int_{z_{\ast}}^{\infty}c_s(z)/E(z)dz$ is the comoving
sound horizon at photo-decoupling epoch.
The redshift $z_{\ast}$ corresponding to the decoupling epoch of photons is given by \cite{CMBz}
\begin{equation}
z_{\ast}=1048[1+0.00124(\Omega_bh^2)^{-0.738}(1+g_{1}(\Omega_{\mathrm{m0}}h^2)^{g_2})],
\end{equation}
where
$g_1=0.0783(\Omega_bh^2)^{-0.238}(1+39.5(\Omega_bh^2)^{-0.763})^{-1}$,
$g_2=0.560(1+21.1(\Omega_bh^2)^{1.81})^{-1}$.
The CMB shift parameter $R$
is expressed as \cite{CMBR}
\begin{equation}
R=\Omega_{\mathrm{m0}}^{1/2}\Omega_\mathrm{k}^{-1/2}sinn\bigg[\Omega_\mathrm{k}^{1/2}\int_0^{z_{\ast}}\frac{dz}{E(z)}\bigg].
\end{equation}

For the CMB data, we choose the data set including the the acoustic
scale ($l_a$), the shift parameter ($R$), and the redshift of
recombination ($z_{\ast}$). The WMAP7 measurement gives the best-fit
values of the data set \cite{CMB}
\begin{eqnarray}
\hspace{-.5cm}\bar{\textbf{P}}_{\rm{CMB}} &=& \left(\begin{array}{c}
{\bar l_a} \\
{\bar R}\\
{\bar z_{\ast}}\end{array}
  \right)=
  \left(\begin{array}{c}
302.09 \pm 0.76\\
1.725\pm 0.018\\
1091.3 \pm 0.91 \end{array}
  \right).
 \end{eqnarray}

The $\chi^2$ value of the CMB observation can be calculated as
\cite{CMB}
\begin{eqnarray}
\chi^2_{\mathrm{CMB}}=\Delta
\textbf{P}_{\mathrm{CMB}}^\mathrm{T}{\bf
C_{\mathrm{CMB}}}^{-1}\Delta\textbf{P}_{\mathrm{CMB}},
\end{eqnarray}
where $\Delta\bf{P_{\mathrm{CMB}}} =
\bf{P_{\mathrm{CMB}}}-\bf{\bar{P}_{\mathrm{CMB}}}$, and the
corresponding inverse  covariance matrix is
\begin{eqnarray}
\hspace{-.5cm} {\bf C_{\mathrm{CMB}}}^{-1}=\left(
\begin{array}{ccc}
2.305 &29.698 &-1.333\\
29.698 &6825.270 &-113.180\\
-1.333 &-113.180 &3.414
\end{array}
\right).
\end{eqnarray}

\begin{figure}[!htbp]
\includegraphics[width=20cm]{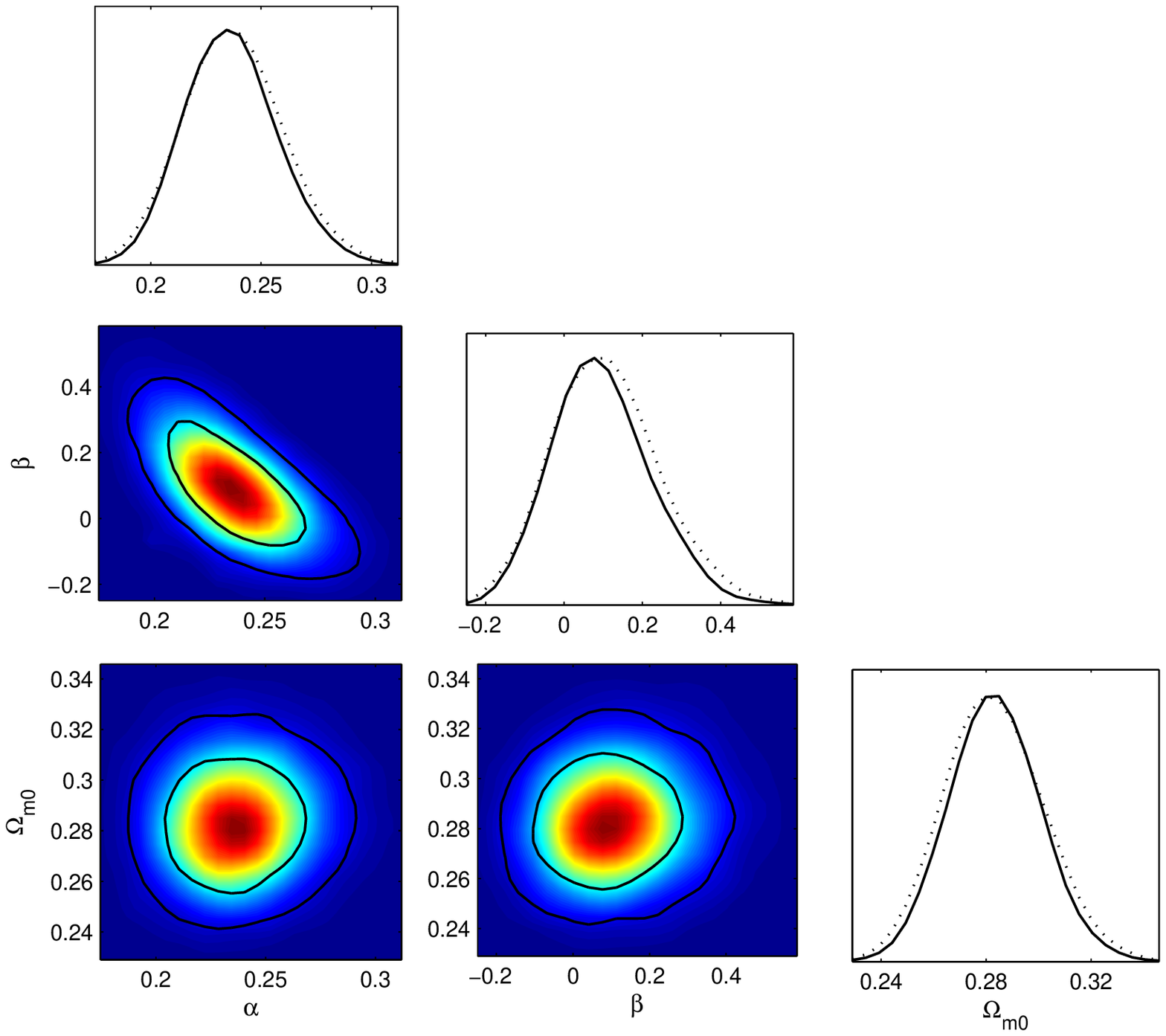}
\caption{The 2-D regions and 1-D marginalized distribution with the
1-$\sigma$ and 2-$\sigma$ contours of parameters
$\Omega_{m0}$, $\alpha$ and $\beta$ in NGR model, for the data sets
SNe+CMB+BAO.
 }\label{figNGR}
\end{figure}

\begin{figure}[!htbp]
\includegraphics[width=20cm]{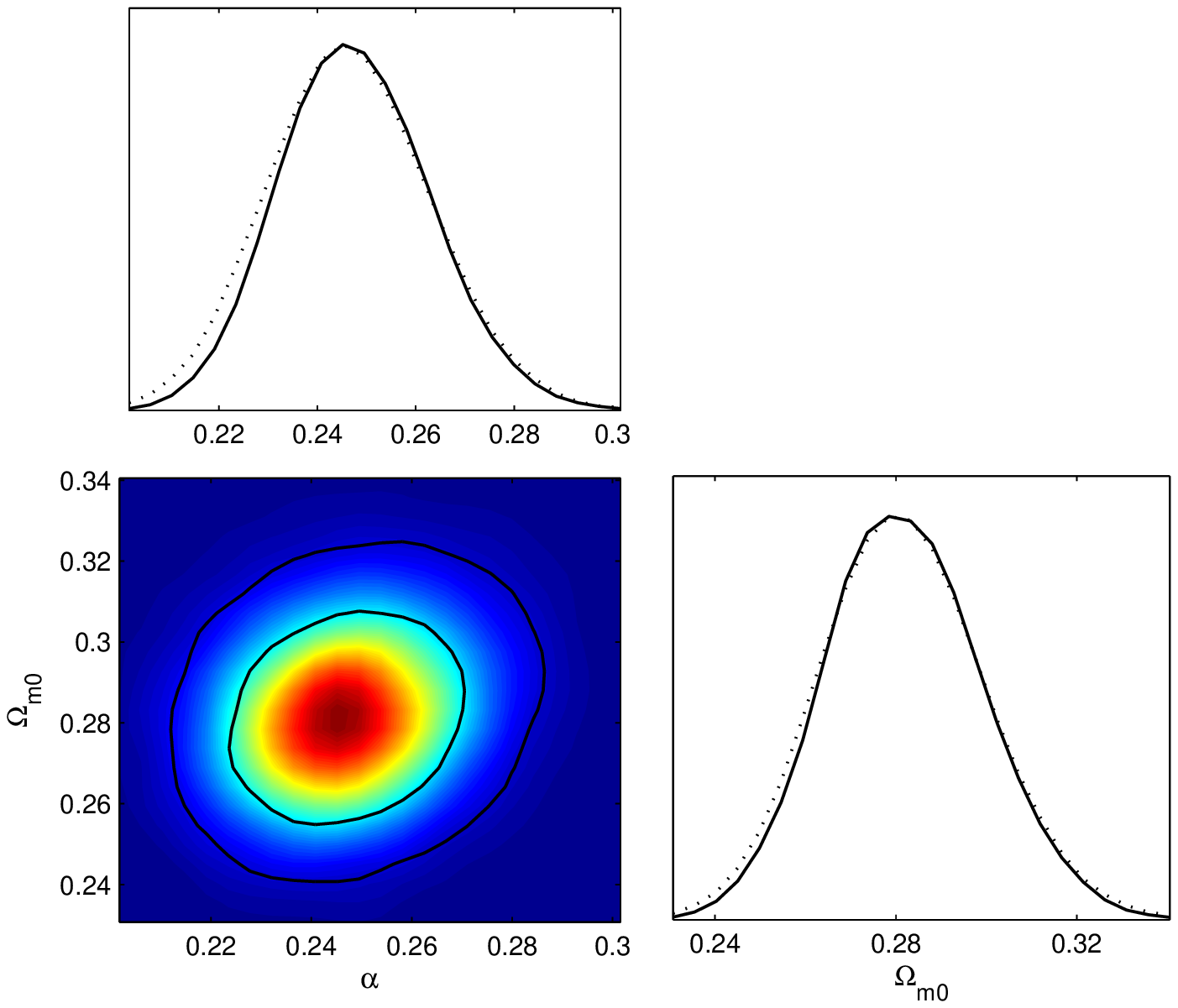}
\caption{The 2-D regions and 1-D marginalized distribution with the
1-$\sigma$ and 2-$\sigma$ contours of parameters
$\Omega_{m0}$ and $\alpha$ in XCDM model, for the data sets
SNe+CMB+BAO.
 }\label{figXCDM}
\end{figure}

\begin{figure}[!htbp]
\includegraphics[width=20cm]{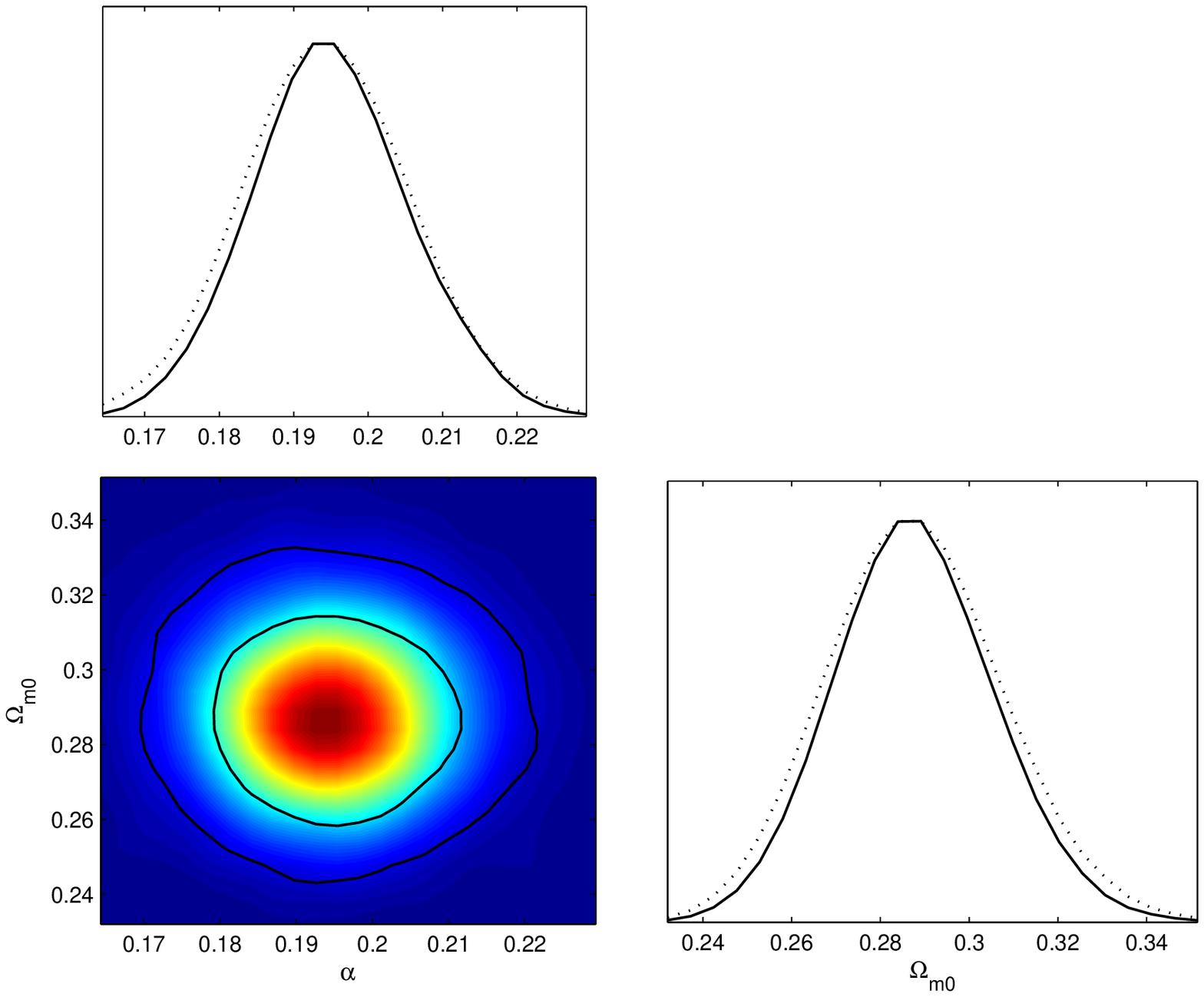}
\caption{The 2-D regions and 1-D marginalized distribution with the
1-$\sigma$ and 2-$\sigma$ contours of parameters
$\Omega_{m0}$ and $\alpha$ in NGR model where we fix $\beta=0.5$, for the data sets
SNe+CMB+BAO.
 }\label{fighalf}
\end{figure}

\begin{figure}[!htbp]
\includegraphics[width=20cm]{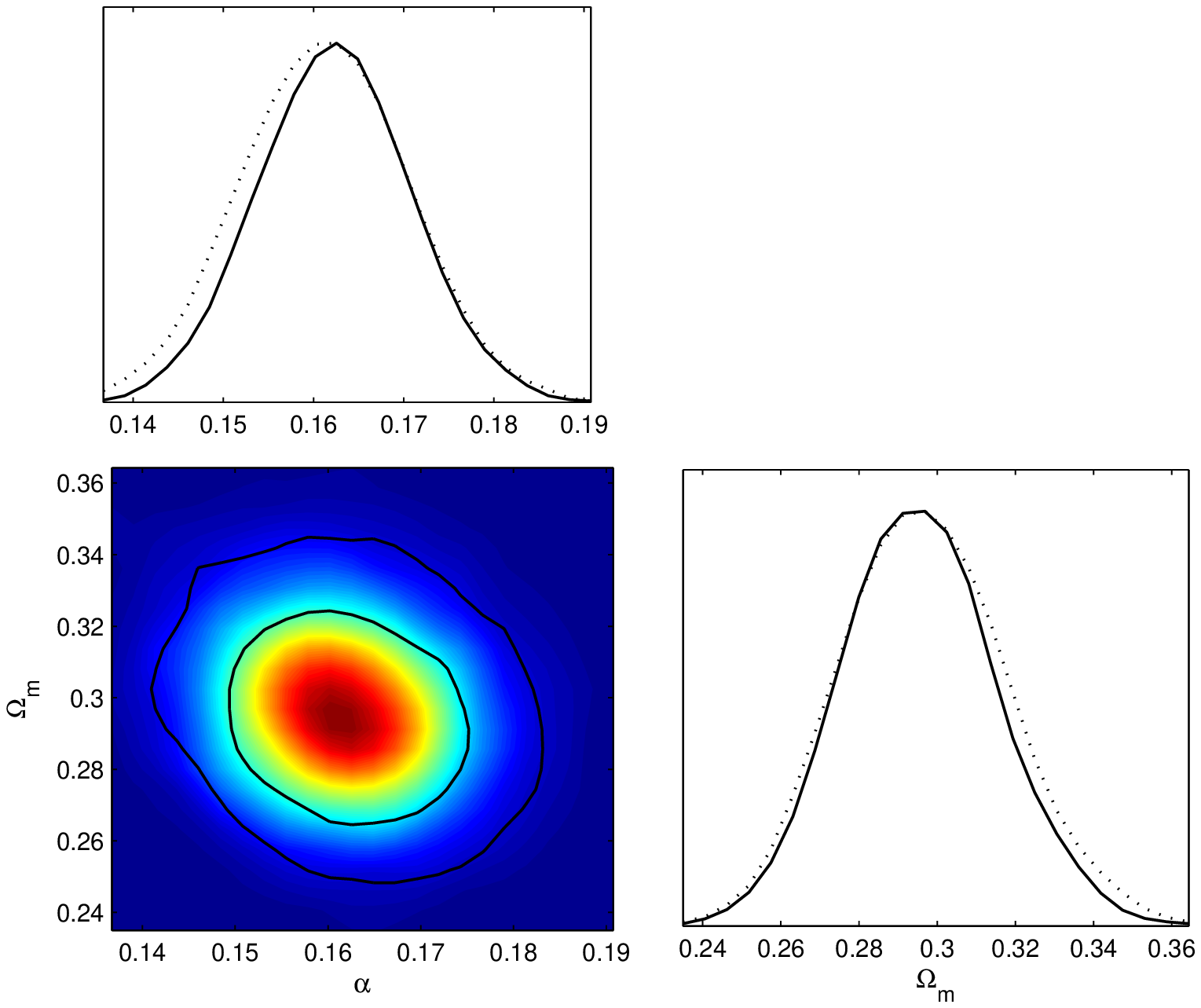}
\caption{The 2-D regions and 1-D marginalized distribution with the
1-$\sigma$ and 2-$\sigma$ contours of parameters
$\Omega_{m0}$ and $\alpha$ in RDE model, for the data sets
SNe+CMB+BAO.
 }\label{figRDE}
\end{figure}

\section{$\text{Constraint results}$}
We choose the common cosmic observations including SNe Ia, BAO and CMB to constrain the NGR model.
We use the usual maximum likelihood method of $\chi^2$ fitting with Markov Chain Monte Carlo (MCMC) method. The code is based on
CosmoMCMC \cite{cosmc}. The total $\chi^{2}$ can be expressed as
\begin{equation}
\chi^2=\chi^2_{\rm SNe}+\chi^2_{\rm BAO}+\chi^2_{\rm CMB}.
\end{equation}

We show the 1-D probability of each parameter ($\Omega_{m0}$, $\alpha$ and $\beta$)
and 2-D plots for parameters between each other for the NGR model in Fig. \ref{figNGR}. The constraint results are
$\Omega_{m0}=0.284_{-0.035}^{+0.036}(1\sigma)_{-0.048}^{+0.050}(2\sigma)$,
$\alpha=0.235_{-0.039}^{+0.046}(1\sigma)_{-0.053}^{+0.068}(2\sigma)$, $\beta$=$0.08_{-0.21}^{+0.30}(1\sigma)_{-0.28}^{+0.43}(2\sigma)$. We can see $\beta=0$ is within
1-$\sigma$ range and $\beta=1$ is ruled out within $2\sigma$ regions.
Moreover, we further fix the value of $\beta$ in three cases: $\beta=0$ (XCDM), $\beta=0.5$ (the situation NGR model permits) and
$\beta=1$ (RDE). The results are plotted in Fig. \ref{figXCDM}, Fig. \ref{fighalf} and Fig. \ref{figRDE}, respectively. Numerical
results are shown in Table. \ref{tab1}. We can see that when $\beta$ becomes larger, the corresponding $\chi_{\rm min}^2$ becomes larger quickly.
The $\chi_{\rm min}^2$ of RDE model is 558.834 while $\chi_{\rm min}^2$ of XCDM model is only 532.238.
We can also see when $\beta$ becomes larger, the density of matter becomes larger and parameter $\alpha$ becomes smaller.
Our constraint results are consistent with \cite{consist}.

\section{$\text{Discussion and conclusion}$}
In this Letter, we propose a new generalized Ricci dark energy model based on the weight of matter. This model contains both
Ricci dark energy model and XCDM model through weight factor $\beta$. $\beta=0$ and $\beta=1$ are corresponding to XCDM model
and RDE model, repectively. Moreover, NGR model permits an arbitrary value of $\beta$. If we fix the EOS parameter today $\omega_{de0}=-1$
and $\Omega_{m0}=0.27$, which seems reasonable for all kind of observations, the larger $\beta$ becomes, the faster EOS parameter
$\omega_{de}$ tends to 0. Besides, deceleration parameter becomes smaller in the future, Hubble parameter becomes larger and density parameter of dark energy becomes larger. The observations can give
us the distribution of $\beta$, which provides a criterion for testing XCDM and RDE. It is similar to testing the
distance-duality relation \cite{DD}. Both of them set the key parameter free. We use the latest observational data including
SNe Ia, BAO and CMB to constrain NGR model. The constraint results tend to supporting XCDM model or even $\Lambda$CDM model (corresponding to $\beta=0$ and $\alpha=0.25$) rather than RDE model.
We can conclude that RDE model is ruled out by the observations we select in NGR scenario within $2\sigma$ regions.
 For future study on this problem, we
hope more data and more independent cosmic methods can give a more confirmed discrimination. We further compare NGR model with some of current successful dark energy
models including Chevallier-Polarski-Linder (CPL) parametrization \cite{CPL}, generalized Chaplygin gas (GCG) and interacting dark energy (IDE) model \cite{IDE} through AIC information criterion. The AIC is defined as $AIC=\chi^2_{min}+2k$, where $k$ is the number of parameters.
We show the comparisons in Table. \ref{tab2}. NGR model as a three-parameter cosmological model can compete with CPL and IDE model.
From the discussions above, we can see NGR model gives a good discrimination between RDE model and XCDM model. Besides, as a unification of RDE model and XCDM model, it can be a good choice for describing the universe itself.

\begin{table}
 \begin{center}
 \begin{tabular}{c c   cc c  } \hline\hline
 Model &             Number of parameters      & $\chi_{\rm min}^2$   & $\Delta AIC$ &  \\ \hline
 $\Lambda$CDM     & 1     &  532.313   &  0   &\\
 XCDM         & 2      & 532.238           &  1.925     & \\
 RDE             &  2         & 558.834         &  28.521   & \\
 GCG               & 2          & 532.159                &  1.846      &  \\
 CPL         & 3  &  531.804      &  3.491   &\\
 IDE              & 3          & 531.712  &  3.399  & \\
 NGR        & 3        & 531.710   & 3.397\\
 \hline\hline
 \end{tabular}
 \caption{The comparisons among various cosmological models through the same method and observations.}\label{tab2}
 \end{center}
 \end{table}

\textbf{\ Acknowledgments } This work was supported by the National Natural Science Foundation of
China under the Distinguished Young Scholar Grant 10825313,
the Ministry of Science and Technology National Basic Science Program (Project 973)
under Grant No.2012CB821804, the Fundamental Research Funds for the Central
Universities and Scientific Research Foundation of Beijing Normal University.

\end{document}